\documentclass{elsart}
\usepackage{graphicx}
\usepackage[dvips]{color}
\definecolor{Black}{named}{Black}
\definecolor{Red}{named}{Red}

\def\gsim{\;\raise0.3ex\hbox{$>$\kern-0.75em\raise-1.1ex\hbox{$\sim$}}\;}
\def\lsim{\;\raise0.3ex\hbox{$<$\kern-0.75em\raise-1.1ex\hbox{$\sim$}}\;}

\def\sw2{\sin^2 \theta_W}
\def\epsilon{\varepsilon}
\def\s132{\sin^2 \theta_{13}}
\newcommand{\neff}{N_{\rm eff}}
\newcommand{\np}{N^\prime}
\def\Tdp{T_d^\prime}
\def\Td{T_d}
\begin{document}
\begin{frontmatter}
{\hfill \small DSF-23/2006, IFIC/06-24, MPP-2006-80}\\
\title{Effects of non-standard neutrino-electron \\
interactions on relic neutrino decoupling}
\author[Napoli]{Gianpiero Mangano},
\author[Napoli]{Gennaro Miele},
\author[IFIC]{Sergio Pastor},
\author[IFIC]{Teguayco Pinto},
\author[Napoli]{Ofelia Pisanti},
\author[MPI]{Pasquale D.\ Serpico}
\address[Napoli]{Dipartimento di Scienze Fisiche, Universit\`{a} di Napoli
{Federico II} and INFN, Sezione di Napoli, Complesso Universitario di
Monte S.\ Angelo\\ Via Cintia, I-80126 Naples, Italy}
\address[IFIC]{Instituto de F\'{\i}sica Corpuscular (CSIC-Universitat de
Val\`{e}ncia),\\
Ed.\ Institutos de Investigaci\'{o}n, Apdo.\ 22085,
E-46071 Valencia, Spain}
\address[MPI]{Max-Planck-Institut f\"{u}r Physik
(Werner-Heisenberg-Institut),\\F\"{o}hringer Ring 6, D-80805 Munich,
Germany}
\begin{abstract}
We consider the decoupling of neutrinos in the early Universe in
presence of non-standard neutral current neutrino-electron
interactions (NSI). We first discuss a semi-analytical approach to
solve the relevant kinetic equations and then present the results of
fully numerical and momentum-dependent calculations, including
flavor neutrino oscillations. We present our results in terms of both
the effective number of neutrino species ($N_{\rm eff}$) and the
impact on the abundance of $^4$He produced during Big Bang
Nucleosynthesis.  We find that the presence of neutrino-electron NSI
may enhance the entropy transfer from electron-positron pairs into
neutrinos instead of photons, up to a value of $N_{\rm eff}\simeq 3.12$
for NSI parameters within the ranges allowed by present laboratory
data, which is almost three times the effect that appears for standard weak
interactions.  Thus non-standard neutrino-electron interactions do not
essentially modify the density of relic neutrinos nor the bounds on
neutrino properties from cosmological observables, such as their mass.
\end{abstract}
\end{frontmatter}
%
\section{Introduction}
\label{sec:introduction}
In the early Universe, neutrinos were kept in thermal contact with the
electromagnetic primordial plasma by rapid weak interactions with
electrons and positrons. When the temperature dropped below a few MeV,
these weak processes became ineffective and the process of neutrino
decoupling took place, while shortly after the $e^\pm$ pairs began to
annihilate almost entirely into photons thus producing a difference
between the temperatures of the relic photons and neutrinos. This
difference can be easily calculated if we assume that neutrinos were
completely decoupled when the $e^\pm$ pairs transferred their entropy
to photons, leading to the well-known temperature ratio
$T_\gamma/T_\nu=(11/4)^{1/3}\simeq 1.40102$. Indeed, this simplified
picture should be improved since some relic interactions between
$e^\pm$ and neutrinos exist all along the $e^\pm$ annihilation stage,
leading to a slightly smaller increase of the comoving photon
temperature and to small distortions (at the percent level) of the
neutrino momentum distributions.

Presently, there exist compelling evidences for flavor neutrino
oscillations from a variety of experimental data on solar,
atmospheric, reactor and accelerator neutrinos (see e.g.\
\cite{Maltoni:2004ei,Fogli:2005cq}). These results are well
understood by assuming that neutrinos have masses and mix, which in
turn seems to point out the necessity of some new physics beyond the
Standard Model (SM) of fundamental interactions.  Interestingly,
non-zero neutrino masses usually come with non-standard interactions
(NSI) that might violate leptonic flavor and/or break weak
universality. Recent analyses
\cite{Berezhiani:2001rs,Berezhiani:2001rt,Davidson:2003ha,Barranco:2005ps}
have considered the neutral current NSI in a phenomenological way,
showing that they can be bound using measurements of
neutrino-electron scattering, as well as data from LEP and from
related charged lepton processes.

The aim of this paper is to study the neutrino decoupling process in
presence of additional interactions between neutrinos and electrons,
a possibility already noted in
\cite{Berezhiani:2001rs,Davidson:2003ha}. In this case, neutrinos
could be kept in longer contact with $e^\pm$ and thus share a larger
amount of the total entropy transfer than in the SM. Actually, if
the non-standard neutrino-electron interactions were large enough,
the neutrino momentum distribution would be significantly different
from the standard case. In turn, this would modify the final yield
of light nuclei during the epoch of Big Bang Nucleosynthesis (BBN),
as well as the radiation content of the Universe, affecting the
anisotropies of the Cosmic Microwave Background (CMB) and the power
spectrum of Large Scale Structures (LSS). Our goal is to calculate
how the decoupling is modified taking into account NSI with
couplings which are still allowed by present laboratory data, and to
discuss the possibility that cosmological observations can be used
as a complementary way to bound these exotic scenarios.

The paper is organized as follows. We begin in Sec.\
\ref{NSIsummary} by describing the formalism adopted for the
non-standard electron-neutrino interactions and summarize the
current bounds from a variety of experimental data. We then consider
the process of relic neutrino decoupling in the presence of
non-standard electron-neutrino interactions, giving first an
estimate by using a semi-analy\-tical approach in Sec.\
\ref{sec:decoupling}. Finally, in Sec.\ \ref{numerical} we report
the results of the full momentum-dependent numerical calculations
for the neutrino spectra and the effect on the primordial $^4$He
yield and other cosmological observables. We present our conclusions in 
Sec.~\ref{conclus}.

\section{Non-standard neutrino-electron interactions}\label{NSIsummary}
The long-standing evidence of flavor change in atmospheric and solar
neutrino experiments represents a strong indication of some new
physics beyond the SM of fundamental interactions. For several years
these experimental results have been interpreted in terms of
neutrino masses (flavor oscillations) or by introducing new neutrino
interactions. Recently, reactor and accelerator data have confirmed
that neutrino masses are indeed the main mechanism explaining the
atmospheric and solar anomalies, while the role of NSI can be only
sub-leading. In any case, most extended particle physics models that
account for neutrino masses naturally lead to new NSI, whose value
strongly depends on the model. For instance, NSI may arise from the
structure of the charged and neutral current weak interactions in
seesaw-type extended models \cite{Schechter:1980gr}.

In the present analysis we will follow
\cite{Berezhiani:2001rs,Davidson:2003ha} and assume that new physics
induces NSI only through the four fermion operators
$(\bar{\nu}\nu)(\bar{f}f)$, where $f$ is a charged lepton or quark,
but not new charged lepton physics at tree level. In particular, since
we are interested in the decoupling process of relic neutrinos, we
consider only the NSI related to electrons which, together with the
standard weak interactions, are described by the effective Lagrangian
\begin{equation}
{\mathcal L}_{\rm eff} = {\mathcal L}_{\rm SM}+
\sum_{\alpha,\beta}{\mathcal L}^{\alpha\beta}_{\rm NSI}
\label{Leff}
\end{equation}
which contains the four-fermion terms
\begin{eqnarray}
{\mathcal L}_{\rm SM} &=& -2\sqrt{2}G_F \,\left \{ (\bar{\nu}_e
\gamma^\mu L\nu_e )(\bar{e} \gamma_\mu L e)  + \sum_{P,\alpha} g_{P}
(\bar{\nu}_\alpha \gamma^\mu L \nu_\alpha)(\bar{e} \gamma_\mu P
e)\right \}
\label{SM}\\
{\mathcal L}^{\alpha\beta}_{\rm NSI} &=& -2\sqrt{2}G_F \,
\sum_{P}\,\epsilon^P_{\alpha\beta}\,
(\bar{\nu}_\alpha \gamma^\mu L
\nu_\beta)(\bar{e} \gamma_\mu P e)
\label{NSI}
\end{eqnarray}
for energies much smaller than the $Z$ boson mass, as in our case,
with $G_F$ the Fermi constant and $P=L,R=(1\mp\gamma_5)/2$ the
chiral projectors. Greek indices label lepton flavors
($\alpha,\beta=e,\mu,\tau$) and the $Z$ couplings are
$g_L=-\frac{1}{2}+\sw2$ and $g_R=\sw2$.

The NSI parameters $\epsilon^P_{\alpha\beta}$ can induce a breaking of
lepton universality ($\alpha=\beta$) or rather a flavor-changing
contribution ($\alpha\neq\beta$). Their values can be constrained by a
variety of laboratory experiments, as discussed in
\cite{Berezhiani:2001rs,Davidson:2003ha,Barranco:2005ps}.
In what follows we summarize the present bounds and refer the reader
to the analyses mentioned above for all details. When considering the
bounds on the $\epsilon^P_{\alpha \beta}$ parameters, it is
important to notice that they are usually obtained taking only
one-at-a-time, or at most combining two of them (such as the pair
$\epsilon^{L,R}_{ee}$). This implies that the derived constraints are
expected to be weaker but more robust if many NSI parameters are
simultaneously included, since cancellations may occur.

A remark is in order. Since isospin-changing weak interactions
converting neutrons into protons and vice versa are important for the
BBN yields, one might wonder if it is legitimate to neglect the
neutrino NSI with quarks. We argue that this approximation is well
justified since, while relic neutrino distributions affect
$n\leftrightarrow p$ processes in a crucial way, vice versa is not
true.  The highly suppressed density of baryons with respect to
electromagnetic particles ($n_b/n_\gamma\simeq 6\times 10^{-10}$)
allows one to neglect completely neutrino scattering on quarks for
calculating the neutrino momentum spectrum.  In addition, what
contributes to the $n\leftrightarrow p$ rates are neutrino
charged-current interactions, where possible non-standard terms are
severely bounded by the accurate agreement of SM calculations for
several processes with the data (e.g.\ leptonic and hadronic weak
decays). For baryon thermalization, neutral current neutrino reactions
are completely negligible with respect to electromagnetic interactions
with $e^\pm$ and photons. Also note that any tiny non-standard effect
in the weak rates is effectively taken into account by our
prescription of rescaling the theoretically predicted neutron lifetime
to the experimentally measured value (see \cite{Serpico:2004gx} for
details).

Finally, we comment on a possible role of exotic four-fermion
neutrino-neutrino interactions of the kind parameterized in Eq.\
(\ref{NSI}), reporting the bounds quoted in \cite{Bilenky:1999dn}. If
only left-handed neutrinos are involved, the accurate measurement of
the $Z-$boson width constrains the NSI coupling to be at most of the
same strength of the neutral current ones in the SM, and thus we
expect that they will have a sub-leading effect on our results (see
Sec.\ \ref{numerical}). Instead, for NSI terms coupling left-handed to
right-handed components, a stringent BBN bound of ${\mathcal
O}(10^{-3})$ applies, coming from the request that more than one
neutrino-equivalent degree of freedom which was thermally populated in
the early Universe is excluded. Finally, extra interactions coupling
only right-handed states might be large, but this case is of no
interest for the process of neutrino decoupling.
\subsection{Bounds from tree level processes}

-- Neutrino scattering experiments

The magnitude of the NSI parameters can be constrained from the
analysis of data from neutrino-electron scattering experiments, which
can probe the SM electroweak predictions with good precision (see
e.g.\ \cite{deGouvea:2006cb}).

\begin{figure}[ht]
\begin{center}
\includegraphics[width=0.7\textwidth,angle=-90]{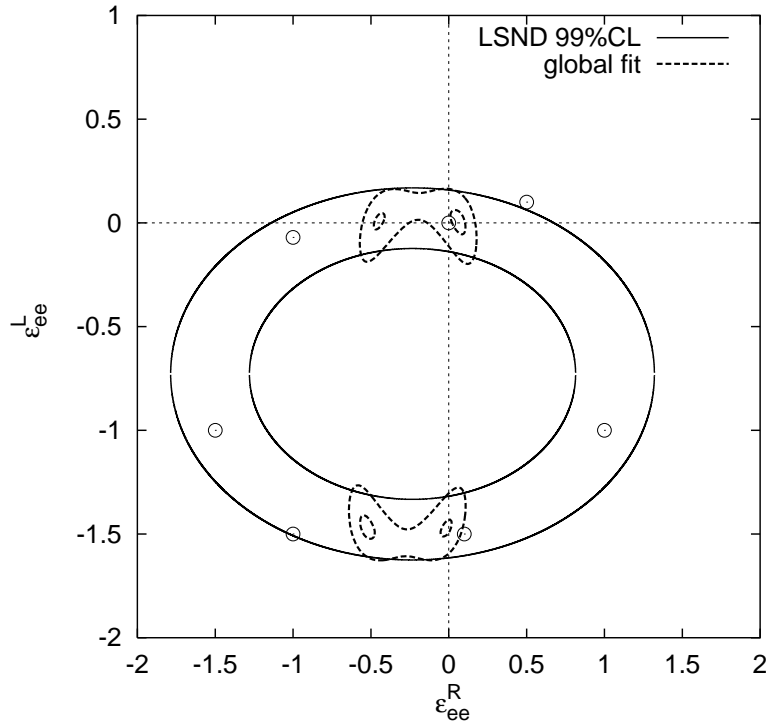}
\caption{The ellipse corresponds to the values of $(\epsilon^{L}_{ee},
\epsilon^{R}_{ee})$ allowed at 99\% CL by LSND data, while the dashed
regions are those allowed at $1\sigma$ and 99\% CL when including data
from antineutrino-electron experiments, as found in
\cite{Barranco:2005ps}.  Here we assumed vanishing off-diagonal NSI
parameters.  We have calculated the full neutrino decoupling process
for the indicated points (see Sec.\ \ref{numerical}).}
\label{epsR_epsL}
\end{center}
\end{figure}

First, we focus on  $\nu_e-e$ (data from the LSND experiment
\cite{Auerbach:2001wg}) and $\bar{\nu}_e-e$ scattering (data from
the Irvine \cite{Reines:1976pv} and MUNU
\cite{Daraktchieva:2003dr} experiments). The total $\nu_e-e$ cross
section including NSI is
\begin{eqnarray}
\sigma(\nu_e  e \rightarrow \nu e) &=& \frac{2G_F^2 m_e
E_\nu}{\pi} \left
[(1+g_L+\epsilon^L_{ee})^2+\frac{1}{3}(g_R+\epsilon^R_{ee})^2
\right. \nonumber \\ &+& \left. \sum_{\alpha \neq e} \left(
\left|\epsilon^L_{\alpha e}\right|^2 + \frac{1}{3} \left|
\epsilon^R_{\alpha e}\right|^2 \right) \right ]
\label{sigmatot_nue_e}
\end{eqnarray}
while the LSND measurement is
\begin{equation}
\sigma(\nu_e  e \rightarrow \nu e) = (1.17 \pm 0.17)
\,\frac{G_F^2 m_e E_\nu}{\pi}  \ ,
\label{sigmatot_LSND}
\end{equation}
In \cite{Davidson:2003ha} this result was compared with the SM
prediction $\sigma=1.0967 G_F^2m_eE_\nu/\pi$---including electroweak
corrections and best fit value for $\sin^2\theta_W$---to obtain the
following allowed regions around the SM value, assuming only one
operator at a time
\begin{eqnarray}
-0.07 \,(-0.14) &<& \epsilon^{L}_{ee} < 0.11\, (0.16)\nonumber\\
-0.99 \,(-1.14) &<& \epsilon^{R}_{ee} < 0.53\, (0.67) \label{bounds_LSND}\\
|\epsilon^{L}_{e \tau}| &<& 0.4 \, (0.5)\nonumber \\
|\epsilon^{R}_{e \tau}| &<& 0.7 \, (0.9)\label{bounds_LSNDfc}
\end{eqnarray}
at 90\% CL (we added 99\% CL bounds in parentheses). For
$\epsilon^{P}_{e \mu}$ more severe bounds are obtained using
radiative effects, see later. In general, the allowed region in the
$\epsilon^{R}_{ee}, \epsilon^{L}_{ee}$ plane is an elliptic corona
as given in Fig.~\ref{epsR_epsL}, assuming $\epsilon^{L,R}_{e
\alpha}=0$ for $\alpha \neq e$.
As shown in \cite{Barranco:2005ps}, with the addition of data from
$\bar{\nu}_e-e$ scattering the allowed LSND region can be
substantially reduced. Since one must exchange $g_L \leftrightarrow
g_R$ with respect to Eq.\ (\ref{sigmatot_nue_e}) one obtains a
perpendicular ellipse. The allowed regions for $\epsilon^{L,R}_{ee}$
are shown in Fig.\ \ref{epsR_epsL}, while the one-parameter 90\% CL
bounds are now \cite{Barranco:2005ps}
\begin{eqnarray}
-0.05\, [-1.58] &<& \epsilon^{L}_{ee} < 0.12\, [0.12]\nonumber\\
-0.04\, [-0.61] &<& \epsilon^{R}_{ee} < 0.14\, [0.15]\nonumber\\
|\epsilon^{L}_{e \tau}| &<& 0.44 \, [0.85]\nonumber \\
|\epsilon^{R}_{e \tau}| &<& 0.27 \, [0.38]\label{bounds_Barranco}
\end{eqnarray}
When leaving the corresponding flavor-changing NSI parameters free,
the bounds are relaxed to the numbers in square parentheses: note the
significant reduction especially for negative $\epsilon^{R}_{ee}$
values.

We also have data on $\nu_\mu-e$ scattering from the CHARM II
collaboration \cite{Vilain:1994qy}. The obtained results for the
vector and axial vector $e-\nu_\mu$ couplings can be translated
into the following 90\% CL bounds on NSI parameters (see e.g.\
\cite{Davidson:2003ha})
\begin{eqnarray}
-0.025 < \epsilon^{L}_{\mu \mu} < 0.03 \nonumber\\
-0.027 < \epsilon^{R}_{\mu \mu} < 0.03 \label{bounds_mu} \\
|\epsilon^{P}_{\mu \tau}| < 0.1 \label{bounds_mufc}
\end{eqnarray}
which imply that the $\nu_\mu-e$ interactions must be very close
to the SM predictions.

-- LEP data on $e^+ e^- \to \nu\bar{\nu}\gamma$

As pointed out in \cite{Berezhiani:2001rs}, neutrino NSI can be also
constrained by measuring the $e^+e^- \to \nu\bar{\nu}\gamma$ cross
section. This is actually the only way to get bounds on the
$\epsilon^{L,R}_{\tau\tau}$ parameters from laboratory data. The
approximate limits for these parameters can be extracted from Fig.~3
of \cite{Berezhiani:2001rt},
\begin{eqnarray}
-0.7 \lsim \epsilon^{L}_{\tau \tau} \lsim 0.5 \nonumber\\
-0.5 \lsim \epsilon^{R}_{\tau \tau} \lsim 0.6 \label{bounds_tau}
\end{eqnarray}
at 99\% CL when the other parameter is left free. For the $\nu_e
\,e$ NSI the bounds are comparable to those obtained from
neutrino-electron scattering.

\subsection{One loop effects}

The non-standard neutrino-electron interactions also lead to
corrections at the one-loop level to processes such as the decays
of the electroweak gauge bosons or lepton flavor violating decays
of charged leptons. These corrections arise from effective
non-renormalizable interactions and in principle their computation
requires the knowledge of the complete theory
leading to these effective terms in the low-energy regime.
Nevertheless for $\Lambda \gg m_W$, where $\Lambda$ is the energy scale
setting the limit of validity of the effective theory, 
the leading term is independent
of the specific theory and has a logarithmic behavior,
$\ln(\Lambda/m_W)$ \cite{Davidson:2003ha}. Using conservatively
$\ln(\Lambda/m_W)\simeq 1$, one gets the following 90\% CL bounds
from the decay rates of the electroweak gauge bosons
\begin{equation}
|\epsilon^{L,R}_{\tau \tau}|\lsim 0.5 \label{bounds_tau2}
\end{equation}
which are of the same order of Eq.\ (\ref{bounds_tau}).

Similarly, from the strong experimental limit on the branching
ratio $Br(\mu^- \rightarrow e^- e^+ e^-) < 10^{-12}$, it is
possible to obtain a severe bound on the flavor changing
parameter (90\% CL)
\begin{equation}
|\epsilon^{P}_{e \mu}| \lsim 5 \times 10^{-4} \, .
\end{equation}

\subsection{Summary of bounds on NSI parameters}

After reviewing the bounds on NSI parameters, we conclude that
they
\begin{itemize}
\item restrict the $\epsilon^{P}_{ee}$ as shown in Fig.\
\ref{epsR_epsL};
\item force the $\epsilon^{P}_{\mu \mu}$ to be very small and
negligible for the purposes of the present paper;
\item allow {\it large} $\epsilon^{P}_{\tau \tau}$, almost of
order unity, see (\ref{bounds_tau});
\item as far as flavor changing parameters are concerned, allow large
$\epsilon^{P}_{e \tau}$, (\ref{bounds_LSNDfc}), moderate
$\epsilon^{P}_{\mu \tau}$, (\ref{bounds_mufc}), and very tiny
$\epsilon^{P}_{e \mu}$, the latter being
too small to have any impact on neutrino decoupling dynamics.
\end{itemize}
Again it is useful to remind the reader that the bounds on NSI parameters
are relaxed when more than one parameter is allowed to vary simultaneously,
as shown in \cite{Barranco:2005ps}.

As noted in \cite{Davidson:2003ha}, the leptonic measurement of
$\sin^2\theta_W$ at a neutrino factory will be sensitive to values of
the $\nu_{e,\mu}-e$ NSI parameters of ${\mathcal O}(10^{-3})$.
Additional information on the NSI can be extracted from a consistency
comparison of the results of solar neutrino experiments with those of
an experiment detecting reactor neutrinos such as KamLAND. The former
depend on the NSI parameters through the matter potential felt by
solar neutrinos and/or via the neutrino neutral current
detection. Instead, KamLAND data is essentially unaffected by the
NSI. Ref.\ \cite{Davidson:2003ha} found that the combination of SNO
and KamLAND data\footnote{The Borexino detector was considered in
\cite{Berezhiani:2001rt}.}  will provide bounds on the
$\epsilon^{P}_{\tau\tau}$ comparable but slightly better to those
in Eqs.\ (\ref{bounds_tau}) or (\ref{bounds_tau2}).
However, it has been noted in several works
\cite{Friedland:2004pp,Guzzo:2004ue,Miranda:2004nb,Friedland:2004ah,Friedland:2005vy}
that while NSI are expected to play a subdominant role on the
oscillations of atmospheric and solar neutrinos, there exist
degenerate directions in the parameter space where large NSI are still
allowed. Future data from short and medium baseline neutrino beams will
help to resolve these degeneracies \cite{Friedland:2006pi}.

\section{Neutrino decoupling in presence of NSI}
\label{sec:decoupling}
\subsection{Delayed neutrino decoupling in the instantaneous limit}
In the early Universe, thermally produced neutrinos were in
equilibrium with other particles down to temperatures of few MeV,
when weak interactions became ineffective and neutrino decoupled
from the plasma. As a first approximation, the
thermal Fermi-Dirac momentum spectrum is preserved after
the freeze-out of weak interactions, since neutrinos decoupled when
ultra-relativistic and both neutrino momenta and temperature
redshift identically with the Universe expansion.
A neutrino chemical potential $\mu_\nu$ would exist in the
presence of a neutrino-antineutrino asymmetry, but it was shown in
\cite{Dolgov:2002ab} that the stringent BBN bounds on
$\mu_{\nu_e}$ (for an updated limit see e.g.\ \cite{Serpico:2005bc})
apply to all flavors, since neutrino oscillations
lead to flavor equilibrium before BBN. Thus we shall ignore
a relic neutrino asymmetry in the following.

It proves useful to define the
following dimensionless variables instead of time, momenta and
photon temperature
\begin{equation}
x \equiv m\,R  \qquad  y \equiv p\,R  \qquad z \equiv T_\gamma\,
R~, \label{comoving}
\end{equation}
where $m$ is an arbitrary mass scale which we choose as the electron
mass $m_e$ and $R$ is the Universe scale factor. The function $R$ is
normalized without loss of generality so that $R(t)\to 1/T$ at large
temperatures, $T$ being the common temperature of the particles in
equilibrium far from any entropy-transfer process. With this choice,
$R^{-1}$ can be identified with the temperature of neutrinos in the
limit of instantaneous decoupling.  After decoupling, neutrinos enter
cosmological observables mainly via their energy density.  Since they
also remain relativistic for most of the cosmological evolution---down
to a temperature of the order of their sub-eV mass scale---it is
customary to parameterize their contribution to the radiation energy
density $\rho_{\rm R}$ in terms of the effective number of neutrinos
$N_{\rm eff}$~\cite{Shvartsman:1969mm,Steigman:1977kc}
\begin{equation}
\neff
\equiv\frac{\rho_{\nu+X}}{\rho_\nu^0}\frac{\rho_\gamma^0}{\rho_\gamma}
=\left(3+\sum_\alpha^3\delta_\alpha\right)\left(\frac{z_0}{z}\right)^4
\label{neff1},
\end{equation}
where $z_0(x)$ describes the photon to neutrino temperature ratio in the limit
of instantaneous neutrino decoupling. For $x\gg 1$ the $e^{\pm}$
annihilation phase is over, and
$z_0\to(11/4)^{1/3}\simeq 1.4010$. The energy densities
$\rho_\gamma^0$ and $\rho_\nu^0$ refer respectively to the photon plasma
and to a single neutrino species in the
limit of instantaneous decoupling, while $\rho_\gamma$ is the actual energy
content of the photon plasma and $\rho_{\nu+X}$ the total energy
content of weakly interacting particles (including possible exotic
contributions). The second equality in Eq.~(\ref{neff1}) follows when
only the three
active neutrinos contribute to $\rho_{\nu+X}$; eventually, the actual
photon temperature evolution accounts for the second factor in Eq.~(\ref{neff1})
and the possible energy-density distortion in the $\alpha$-th neutrino
flavor is given by
$\delta_\alpha\equiv(\rho_{\nu_\alpha}-\rho_\nu^0)/\rho_\nu^0$.
Note that from Eq.~(\ref{neff1}) it follows
\begin{equation}
\rho_{\rm R} = \left( 1 + \frac{7}{8}z_0^4 \, N_{\rm eff} \right) \,
\rho_\gamma \,\,, \label{neff}
\end{equation}
which, replacing $z_0$ with $(11/4)^{4/3}$, is often used in the literature
to define $\neff$ in the asymptotic limit $x\gg 1$.
Clearly, well after $e^\pm$ annihilation, three thermally distributed neutrinos
correspond to $N_{\rm eff}=3$ in the instantaneous decoupling limit.

In order to estimate analytically how large would be the impact of NSI
on the decoupling mechanism, let us discuss a simple toy-model, which
we shall later compare with the results obtained solving the relevant
kinetic equations.  Consider the general case where $3-\np$ neutrinos
interact via standard weak processes, while the remaining $\np$ have
extra contributions from NSI which enhance the interaction rates over
$e^\pm$. In the instantaneous decoupling limit for both the neutrino
species, the plasma can be described by a single additional parameter,
the non-standard neutrino temperature $T^\prime$ or, equivalently, by
$w\equiv T^\prime\, R\geq 1$.  From Eq.~(\ref{neff1}), one obtains
\begin{equation}
N_{\rm eff}=\left[(3-\np)+\np\,w^4 \right]\left(\frac{z_0}{z}\right)^4
=\left[3+\np(w^4-1) \right]\left(\frac{z_0}{z}\right)^4,
\label{neff-entropy}
\end{equation}
where the second step reflects the fact that, for a single NSI
neutrino species, the distortion is given by $\delta^\prime=(w^4-1)$.
If we denote with $x_d\equiv T_d R$ the epoch of the standard neutrino
decoupling, and parameterize with $x_d^\prime\equiv \Tdp R$ the
decoupling temperature of the non-standard species, we can easily
calculate the functions $w(x)$ and $z(x)$ from the conservation of
entropy per comoving volume,
\begin{displaymath}
\left\{
\begin{array}{l}
w(x)=z(x)=1\:\:\:\:\:\:\:\:\:\:\:\:\:\:\:\:\:\:\:\:\:\:\:\:
\:\:\:\:\:\:\:\:\:\:\:\:\:\:\:\:\:\:\:\:\:\:\:\: x< x_d,\\
w(x)=z(x)=\left[
\frac{\np s_\nu(x_d)+s_\gamma(x_d)+s_{e^\pm}(x_d)}
{\np s_\nu(x)+s_\gamma(x)+s_{e^\pm}(x)}\right]^{1/3}\:\: x_d\leq x\leq x_d^\prime,\\
\left.\begin{array}{l}
w(x)=w_d\equiv\left[\frac{\np s_\nu(x_d)+s_\gamma(x_d)+s_{e^\pm}(x_d)}
{\np s_\nu(x_d^\prime)+s_\gamma(x_d^\prime)+s_{e^\pm}(x_d^\prime)}\right]^{1/3} \\
z(x)=w_d\left[
\frac{s_\gamma(x_d^\prime)+s_{e^\pm}(x_d^\prime)}
{s_\gamma(x)+s_{e^\pm}(x)}\right]^{1/3}\\
\end{array}\right\}
\:x> x_d^\prime,
\end{array}
\right.
\end{displaymath}
where $s_\nu$,
$s_\gamma$ and $s_{e^\pm}$ are respectively the specific entropy
of one $\nu$--$\bar{\nu}$ species, of the photons and of the
$e^\pm$.

\begin{figure}[t]
\begin{center}
\includegraphics[width=0.7\textwidth,angle=-90]{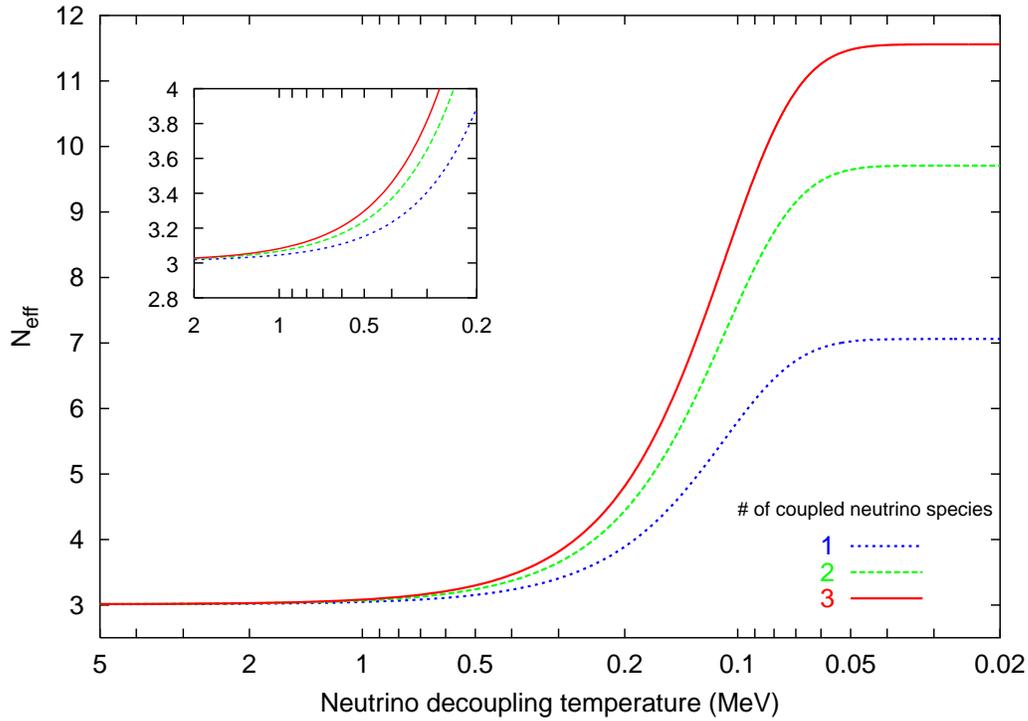}
\caption{Contribution to the radiation energy density parameterized
in terms of $N_{\rm eff}$, when 1, 2 or 3 neutrino species are
coupled to the electromagnetic plasma until a given value of the
decoupling temperature $\Tdp$.} \label{fig_Neff}
\end{center}
\end{figure}
In Fig.~\ref{fig_Neff} we show the results for $\neff$ in the instantaneous
decoupling limit as function of the decoupling temperature $\Tdp$ and for
various choices of $\np$. As expected, $\neff$
grows if neutrinos decouple at a smaller temperature and shows
two regimes. For very small NSI contribution to the neutrino
interaction rates we have $w_d\simeq 1$ and $\neff\simeq 3$
since neutrinos decouple long before any $e^\pm$ annihilation.
On the other hand, for $\np$ neutrinos tightly coupled to the electromagnetic
component down to very small temperatures when all $e^\pm$
pairs have already annihilated,
\begin{equation}
z \simeq \left(\frac{22+7 \np}{8+7\np}\right)^{1/3}, \,\,\,\,\,\,\,
w\simeq z .\label{zf_fully_c}
\end{equation}
Thus, in the tightly coupled limit, one finds
$\neff\to \{7.0, 9.7,11.5\}$ for $\np=\{1,2,3\}$
respectively, as shown in Fig.\ \ref{fig_Neff}.
Note that one obtains a contribution to the radiation energy
density of order $N_{\rm eff}\sim 4$ if neutrinos are kept in
equilibrium down to temperatures of $0.2-0.3$ MeV.

\subsection{Decoupling temperature with NSI}\label{decT_NSI}
In order to dynamically predict the decoupling temperature $\Tdp$ when
NSI are present, we have to solve the relevant kinetic equations
taking into account the new couplings $\epsilon$.  For a first
analytical estimate, we can compare the interaction rate $\Gamma$ with
$e^\pm$ with the expansion rate of the Universe given by the Hubble
parameter $H$.  Since $H\propto T^2$, and $\Gamma\sim (1+\epsilon)^2
T^5$, the decoupling temperature at which $H=\Gamma$ decreases as
$\Tdp\sim (1+\epsilon)^{-2/3}$.  As we have argued in the previous
section, in order to produce changes of ${\mathcal O}(1)$ in $\neff$
the decoupling temperature should be lowered down to
$\Tdp\sim$0.2--0.3 MeV, i.e. should be one order of magnitude smaller
than for ordinary neutrinos.  This implies $\epsilon\gsim 20$,
which would largely exceed present laboratory bounds. We thus expect
at most changes in $\neff$ of ${\mathcal O}$(0.01--0.1). To improve
further our treatment, we can use an approximate solution of the
kinetic equation for the neutrino distribution function $f_\nu(x,y)$,
\begin{equation}
Hx\frac{\partial f_\nu(x,y)}{\partial x}=I_\nu^{\rm coll}
\label{kin1}
\end{equation}
where $I_\nu^{\rm coll}$ is the collision term that includes all
relevant neutrino interaction processes. Following~\cite{Dolgov:2002wy},
we shall keep only the direct reaction term
in the collision integral using the appropriate matrix elements
for neutrino elastic scattering and the inverse annihilation
$e^+e^-\to\bar{\nu}\nu$ in the relativistic limit ($m_e\to 0$).
Assuming that all particles that interact are
close to equilibrium, the collision integral can be estimated in
the Boltzmann approximation leading to
\begin{equation}
Hx\frac{\partial f_\nu}{f_\nu\partial x}\simeq
-\frac{80G^2_F(g^2_L+g^2_R)y}{3\pi^3x^5}
\label{dolg_estim}
\end{equation}
with $g^2_L\to (1+g_L)^2$ for the case of $\nu_e$.  This equation can be
easily integrated in $x$, leading to the values
$\Td(\nu_e)\simeq 1.8$ MeV and $\Td(\nu_{\mu,\tau})\simeq 3.1$ MeV
for an average momentum of $\langle y\rangle=3.15$.
\begin{figure}[t]
\begin{center}
\includegraphics[width=0.48\textwidth]{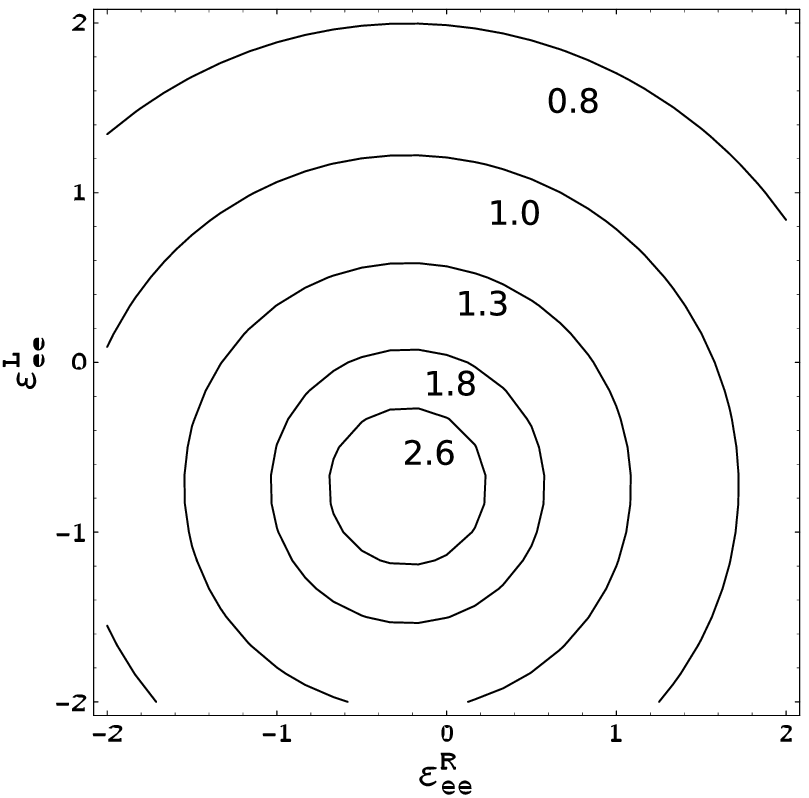}
\includegraphics[width=0.48\textwidth]{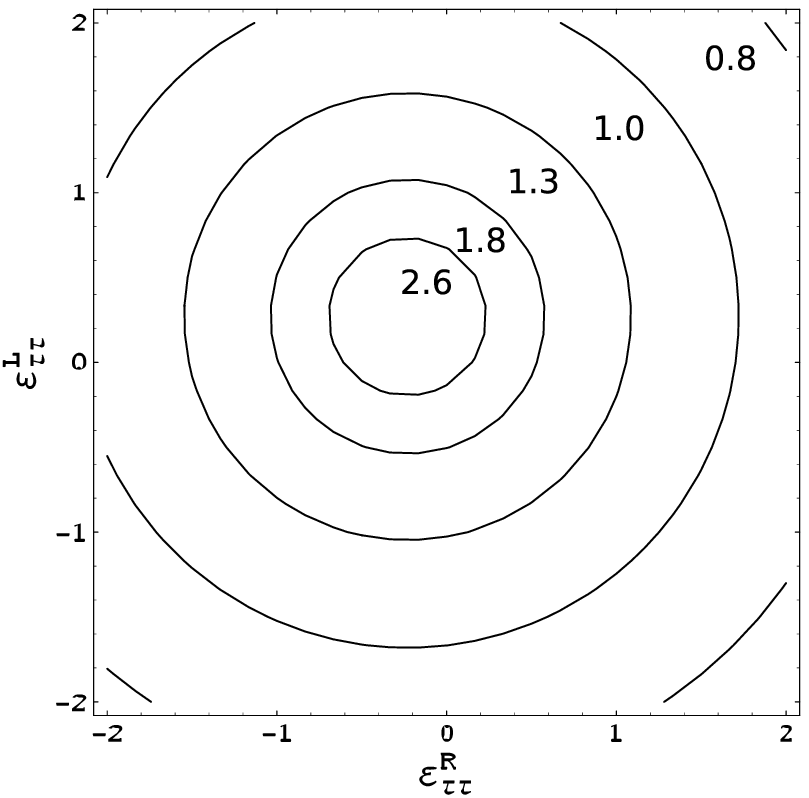}
\caption{Contours of equal neutrino decoupling temperature in MeV for
different values of the parameters characterizing the $\nu_e-e$ and
the $\nu_\tau-e$ diagonal NSI.}
\label{extremecontour}
\end{center}
\end{figure}
This exercise can be repeated in the presence of neutrino-electron
NSI in order to find their influence over the decoupling of
neutrinos, adding $\epsilon^{L,R}_{ee}$ or $\epsilon^{L,R}_{\tau\tau}$ to
the couplings $g_L$ and $g_R$. The calculated decoupling
temperature is shown in Fig.~\ref{extremecontour} for the
case of non-zero $\epsilon^{L,R}_{ee}$ and
 $\epsilon^{L,R}_{\tau\tau}$. In general, a significant increase
of the NSI parameters from the SM prediction leads to a larger
interaction of neutrinos with $e^\pm$ and thus to a lower decoupling
temperature. However, for a region close to the pair of values
$(\epsilon^L,\epsilon^R)$ that minimize the interaction by
accidental cancellation with SM couplings, the decoupling
temperature is significantly raised. Finally, one can relate the
value of the decoupling temperature in presence of NSI with an
estimate of the change in $N_{\rm eff}$ following the instantaneous
decoupling approximation previously described. The results are shown
in Fig.~\ref{eps_neff}. As already noticed, large modifications of
$N_{\rm eff}$ require that the neutrino-electron interactions are
much larger than ordinary weak processes and are thus excluded by
laboratory bounds.

Nonetheless, this study suggests that additional distortions in the
neutrino spectra comparable to or larger than the ones predicted in
the SM are possible, and it is still interesting to assess exactly
how large these effects might be. However, the treatment followed
till now assumes a series of approximations (such as $m_e\to 0$, no
QED corrections to the plasma, no effect of the active neutrino
oscillations, etc.)  that in the next section we shall check using a
fully numerical and momentum-dependent calculation.
\begin{figure}[t]
\begin{center}
\vspace{-2cm}
\hspace{-1.5cm}
\includegraphics[width=0.92\textwidth]{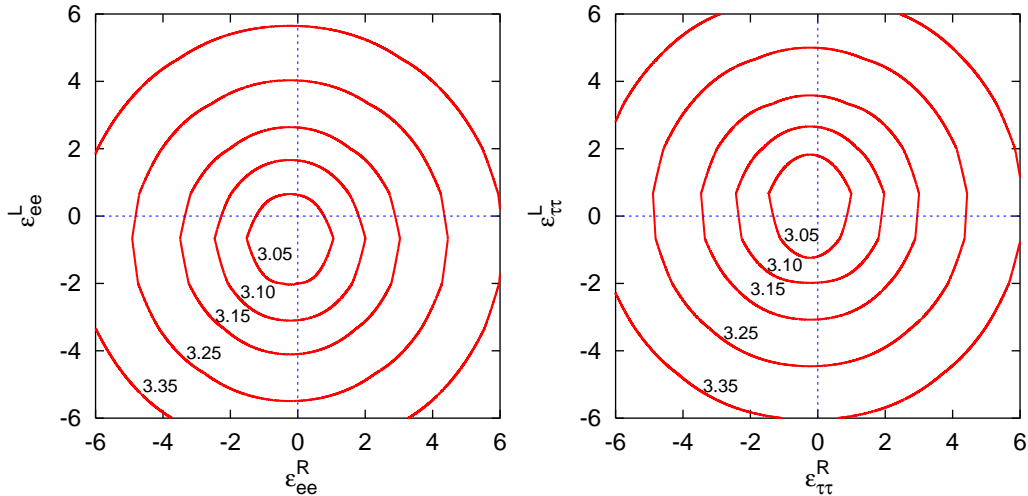}
\caption{Contours of equal $N_{\rm eff}$ for different values of the
NSI parameters as found in the instantaneous decoupling
approximation.}
\label{eps_neff}
\end{center}
\end{figure}

\section{Numerical calculation: momentum-dependent kinetic equations}
\label{numerical}

\subsection{Method}

The instantaneous decoupling approximation does not take into
account that neutrino-electron interactions are more efficient for
larger neutrino energies. Thus one expects a priori that more
energetic neutrinos will remain longer in thermal contact, leading
in general to non-thermal distortions in the neutrino spectra and
a slightly smaller increase of the comoving photon temperature
$z$. This is actually the case for standard weak interactions, as
noted in previous works (for early references, see
\cite{Dicus:1982bz} and the full list given in the review
\cite{Dolgov:2002wy}).

A proper calculation of the process of non-instantaneous neutrino
decoupling demands solving the momentum-dependent Boltzmann equations
for the neutrino spectra, a set of integro-differential kinetic
equations that are difficult to solve numerically. In the early 1990s
several works \cite{Dodelson:1992km,Dolgov:1992qg,Fields:1992zb}
performed momentum-dependent calculations assuming some
approximations, such as Boltzmann statistics for neutrinos, while the
full numerical computation was later carried out
in~\cite{Hannestad:1995rs,Dolgov:1997mb,Dolgov:1998sf,Esposito:2000hi}.
Finally, a further refinement involves the inclusion of finite
temperature QED corrections to the electromagnetic plasma,
\cite{Fornengo:1997wa,Mangano:2001iu}, while the last works also
include the effect of flavor neutrino oscillations
\cite{Hannestad:2001iy,Mangano:2005cc}.

Here we will follow our previous work~\cite{Mangano:2005cc}. In
particular, we will describe the neutrino ensemble in the usual way by
generalized occupation numbers, i.e.\ by $3 {\times} 3$ density
matrices $\varrho$ for neutrinos and anti-neutrinos as described
in~\cite{Sigl:1993fn,McKellar:1992ja}, with elements
$\varrho_{\alpha\beta}$ where $\alpha,\beta=e,\mu,\tau$. The diagonal
elements correspond to the usual occupation numbers of the different
flavors, while the off-diagonal terms are non-zero in the presence of
neutrino mixing or flavor-changing NSI. The equations of motion for
the density matrices relevant for our situation of interest in an
expanding Universe are \cite{Sigl:1993fn}
\begin{equation}\label{eq:3by3evol}
i\left(\partial_t-Hp\,\partial_p\right)\varrho_p= \left[ \left(
\frac{M^2}{2p} -\frac{8 \sqrt2 G_{\rm F}\,p}{3 m_{\rm W}^2}{E}
\right),\varrho_p\right] +{C}[\varrho_p]~,
\end{equation}
where $m_{\rm W}$ is the $W$ boson mass. We use the notation
$\varrho_p=\varrho(p,t)$ and $[{\cdot},{\cdot}]$ denotes the
commutator. The vacuum oscillation term is proportional to $M^2$,
the mass-squared matrix in the flavor basis that is related to the
diagonal one in the mass basis ${\rm diag}(m_1^2,m_2^2,m_3^2)$ via
the neutrino mixing matrix $U$, which in turn depends on the
neutrino mixing angles $\theta_{12}, \theta_{23}$ and
$\theta_{13}$ (we assume CP conservation). From a global analysis
of experimental data on flavor neutrino oscillations, the values
of mixing parameters can be extracted.  As a reference, we take
the best-fit values from~\cite{Maltoni:2004ei}
\begin{equation}
\left(\frac{\Delta m^2_{21}}{10^{-5}~{\rm eV}^2},
\frac{\Delta m^2_{31}}{10^{-3}~{\rm eV}^2},{\rm s}^2_{12},
{\rm s}^2_{23}, {\rm s}^2_{13}\right)= (8.1, 2.2, 0.3,0.5,0)
\label{oscpardef}
\end{equation}
where ${\rm s}^2_{ij}=\sin^2 \theta_{ij}$ for $i,j=1,2,3$.
In Ref.~\cite{Mangano:2005cc} we have shown that the results are not
essentially modified for other values within the allowed regions, in
particular for non-zero $\theta_{13}$ close to the experimental upper
bound. Since we assume maximal $\theta_{23}$ and zero $\theta_{13}$
the spectra of $\nu_\mu$ and $\nu_\tau$ will be the same after
decoupling.

The effects of the medium in Eq.\ (\ref{eq:3by3evol}) are contained in
the collision term $C[{\cdot}]$ and in a refractive term that
corresponds in Eq.\ (\ref{eq:3by3evol}) to the term proportional to
the diagonal matrix $E$, that represents the energy densities of
charged leptons. See \cite{Mangano:2005cc} for a description of both
terms.
The kinetic equations (\ref{eq:3by3evol}) should be numerically
solved along with the equation governing the evolution of the
electromagnetic temperature during the process of $e^\pm$
annihilations. This can be found from the continuity equation for
the total energy density $\rho$,
\begin{equation}
\frac{d\rho}{dt} = -3H\,(\rho+P) \,\,, \label{energy}
\end{equation}
with $P$ the pressure, which can be cast into a first order
differential equation for $z$ as function of $x$, see e.g.\
\cite{Mangano:2001iu}.

We have generalized the results of \cite{Mangano:2005cc} calculating
the full evolution of neutrino during decoupling in presence of
non-standard interactions with electrons, which modify the equations
in two different ways.  First, the couplings are modified inserting
the parameters $\epsilon^{L,R}_{ee}$, $\epsilon^{L,R}_{\tau\tau}$ or
$\epsilon^{L,R}_{e\tau}$ , which implies a variation on the
collisional term in Eq.\ (\ref{eq:3by3evol}). Second, neutrino
refraction in the medium is modified by the NSI, which we took into
account in the refractive term in Eq.\ (\ref{eq:3by3evol})
\begin{equation}
H_{\rm matt}=-\frac{8\sqrt{2}G_Fp}{3m^2_W}\rho_e
\left(\matrix{
1+\epsilon_{ee} & 0 & \epsilon_{e\tau}\cr
0 & 0
& 0\cr
\epsilon_{e\tau} & 0 & \epsilon_{\tau\tau}}
\right)
\end{equation}
where $\rho_e$ is the energy density of the background electrons and
positrons and
$\epsilon_{\alpha\beta}=\epsilon^L_{\alpha\beta}+\epsilon^R_{\alpha\beta}$,
since matter effects are sensitive only to the vector component of
the interaction. The above equation includes only those NSI
parameters that are not constrained to be smaller than ${\mathcal
O}$(0.1) (see Sec.\ \ref{NSIsummary}), and thus represents the
leading-order correction to the refractive index.

We have numerically solved the kinetic equations Eq.\
(\ref{eq:3by3evol}) using a discretization in a grid of $100$
dimensionless neutrino momenta in the range $y_i \in [0.02,20]$. We
start to compute the evolution of the system at a value $x_{\rm
in}$, when weak interactions were effective enough to keep neutrinos
in equilibrium with the electromagnetic plasma but flavor
oscillations are suppressed by medium effects. The system of
equations is then solved from $x_{\rm in}$ until a value of $x$ when
both the neutrino distortions and the comoving photon temperature
$z$ are frozen, approximately at $x_{\rm fin}\simeq 35$. In some
cases, we made the calculation for different initial conditions and
concluded that the results are stable within per cent accuracy 
unless we start the numerical evaluation at $x_{\rm in}\gsim 0.2$.
For other technical details, we refer the reader to~\cite{Mangano:2005cc}.

\subsection{Numerical results}

We have numerically calculated the evolution of the neutrino density
matrix solving the system of Eqs.\ (\ref{eq:3by3evol}) and
(\ref{energy}), during the full process of neutrino decoupling.
Since a complete scan of all possible combinations of the NSI
parameters can not be done, we have performed the calculation for a
selection of values for the NSI parameters, representative of the
regions allowed by present bounds (see Sec.\ \ref{NSIsummary}).
\begin{figure}[t]
\includegraphics[width=.95\textwidth]{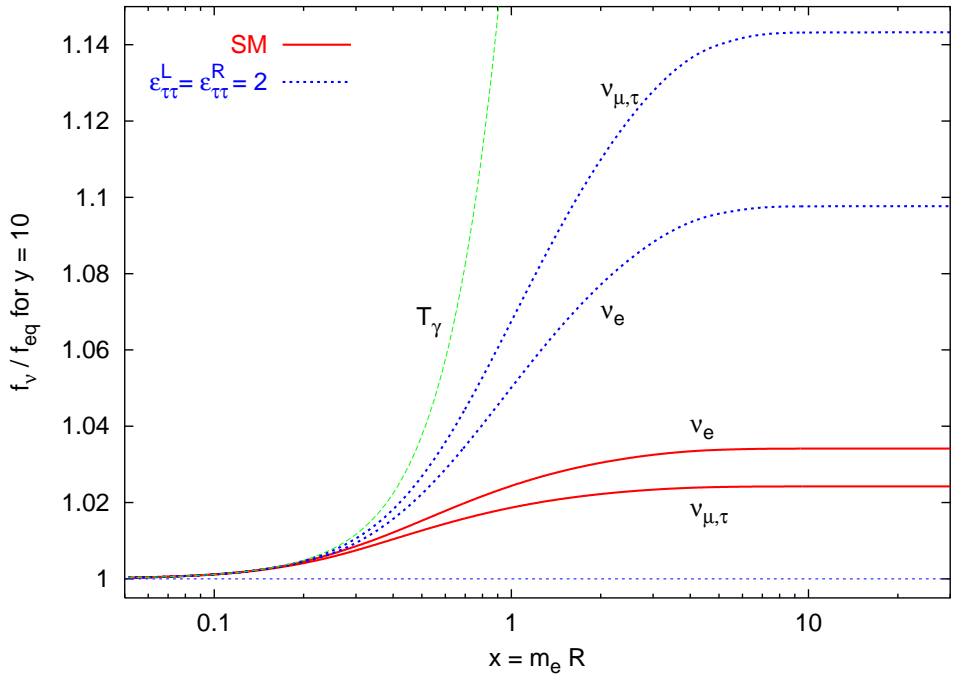}
\caption{\label{fig:evolfnu} Evolution of the distortion of the
$\nu_e$ and $\nu_{\mu,\tau}$ spectra for a particular comoving
momentum ($y=10$) with standard weak interactions (solid line) and
with $\nu_\tau-e$ NSI (dotted line). The line labelled with
$T_\gamma$ corresponds to the distribution of a neutrino in full
thermal contact with the electromagnetic plasma.}
\end{figure}

As an example, in Fig.~\ref{fig:evolfnu} we plot the distortion of the
neutrino distribution as a function of $x$ for a particular neutrino
momentum ($y=10$), both in the case of SM weak interactions and in
presence of large NSI between tau neutrinos and $e^\pm$, corresponding
to $\epsilon^L_{\tau\tau}=\epsilon^R_{\tau\tau}=2$. The behavior of
$f_\nu$ has been previously described e.g.\ in
\cite{Dolgov:1997mb,Esposito:2000hi,Mangano:2005cc}. At large
temperatures or $x\lsim 0.2$, neutrinos are in good thermal contact
with $e^{\pm}$ and their distributions only change keeping an
equilibrium shape with the photon temperature $[\exp(y/z(x))+1]^{-1}$
(the $T_\gamma$ line in the figure). In the intermediate region $0.2
\lsim x \lsim 4$, the standard weak interactions become less effective
in a momentum-dependent way, leading to distortions in the neutrino
spectra which are larger for $\nu_e$'s than for the other
flavors. Finally, at larger values of $x$ neutrino decoupling is
complete and the distortions reach their asymptotic values.  In the
case of large $\nu_\tau-e$ NSI, neutrinos are kept in thermal contact
with $e^{\pm}$ for a longer time, leading to larger distortions for
all neutrino flavors, in particular for $\nu_{\mu,\tau}$ (although
$\epsilon^{L,R}_{ee}=\epsilon^{L,R}_{\mu\mu}=0$, the electron and muon
neutrinos also acquire significant distortions due to the effect of
flavor oscillations). For the particular neutrino momentum in Fig.\
\ref{fig:evolfnu}, if $\epsilon^L_{\tau\tau}=\epsilon^R_{\tau\tau}=2$
the final values of the distribution are $9.8\%$ for the $\nu_e$'s and
$14.3\%$ for the $\nu_{\mu,\tau}$'s larger than in the limit of
neutrino decoupling before any $\e^\pm$ annihilation. For comparison,
the corresponding values for standard weak interactions are $4.4\%$
for the $\nu_e$'s and $2\%$ for the $\nu_{\mu,\tau}$'s.

As in Ref.~\cite{Mangano:2005cc}, we will summarize the results in
terms of the following frozen values: dimensionless photon temperature
$z_{\rm fin}$, the fractional changes in the neutrino energy densities
$\delta_\alpha$ and the asymptotic effective number of
neutrinos $N_{\rm eff}$ as defined in Eq.~(\ref{neff1}).
We have considered four main sets of NSI parameters:

(1) Only $\epsilon^{L,R}_{ee}$: our results are summarized in Table
1. We have calculated various combinations of the two $\nu_e-e$ NSI
parameters, that correspond to the points shown in Fig.\
\ref{epsR_epsL}. For illustrative purposes, we considered the case
with extremely large NSI $\epsilon^L_{ee}=\epsilon^R_{ee}=4$
(already excluded by laboratory bounds), where the distortions in
the distribution functions of neutrinos lead to a significant change
in $N_{\rm eff}$.
\begin{table*}
\begin{center}
\caption{Frozen values of $z_{\rm fin}$, the neutrino energy densities
distortion $\delta_\alpha$, $N_{\rm eff}$ and $\Delta Y_p$ in presence
of $\nu_e-e$ NSI.}
\begin{tabular}{ccccccc}
\hline
$\epsilon^L_{ee}$ & $\epsilon^R_{ee}$ &
$z_{\rm fin}$ &
$\delta_e(\%)$ &
$\delta_{\mu,\tau}(\%)$ &
$N_{\rm eff}$  & $\Delta Y_p$\\ \hline
0 & 0 & 1.3978 & 0.73 & 0.52 & 3.046 & $2.1 {\times} 10^{-4}$\\
\hline
 0.1  &  0.5  & 1.3969 & 1.20 & 0.72 & 3.062 & 2.8${\times} 10^{-4}$\\
-1.0  &  1.0  & 1.3966 & 1.32 & 0.77 & 3.067 & 3.0${\times} 10^{-4}$\\
-1.5  &  0.1  & 1.3976 & 0.81 & 0.56 & 3.048 & 2.0${\times} 10^{-4}$\\
-1.5  & -1.0  & 1.3969 & 1.17 & 0.71 & 3.061 & 2.7${\times} 10^{-4}$\\
-1.0  & -1.5  & 1.3977 & 1.45 & 0.83 & 3.060 & 2.9${\times} 10^{-4}$\\
-0.07 & -1.0  & 1.3973 & 0.99 & 0.64 & 3.055 & 2.6${\times} 10^{-4}$\\
\hline
4.0 & 4.0 & 1.3812 & 9.47 & 3.83 & 3.357 & 11.7${\times} 10^{-4}$\\
\hline\label{tab:eps_e}
\end{tabular}
\end{center}
\end{table*}

(2) Only $\epsilon^{L,R}_{\tau\tau}$: our results are summarized in
Table 2 for values of both parameters up to order unity in absolute
value. Again, we also show one case with larger NSI
$\epsilon^L_{\tau\tau}=\epsilon^R_{\tau\tau}=2$, disfavored by
terrestrial experiments.
\begin{table*}
\begin{center}
\caption{Same as Table \ref{tab:eps_e} but for $\nu_\tau-e$ NSI.}
\begin{tabular}{ccccccc}
\hline
$\epsilon^L_{\tau\tau}$ & $\epsilon^R_{\tau\tau}$ &
$z_{\rm fin}$ &
$\delta_e(\%)$ &
$\delta_{\mu,\tau}(\%)$ &
$N_{\rm eff}$  & $\Delta Y_p$\\ \hline
0 & 0 & 1.3978 & 0.73 & 0.52 & 3.046 & $2.1 {\times} 10^{-4}$\\
\hline
0.5 & 0.5 & 1.3970 & 0.90 & 0.79 & 3.059 & 4.0${\times} 10^{-4}$\\
1.0 & 1.0 & 1.3966 & 1.24 & 1.43 & 3.079 & 4.8${\times} 10^{-4}$\\
-1.0 & -1.0 & 1.3952 & 1.24 & 1.45 & 3.092 & 5.3${\times} 10^{-4}$\\
\hline
2.0 & 2.0 & 1.3911 & 2.04 & 2.96 & 3.168 & 10.0${\times} 10^{-4}$\\
\hline\label{tab:eps_tau}
\end{tabular}
\end{center}

\end{table*}

(3) Combinations of $\epsilon^{L,R}_{ee}$ and $\epsilon^{L,R}_{\tau\tau}$: our
results are summarized in Table 3. Here we varied the values of the
$\epsilon^{L,R}_{ee}$ pair, while the $\nu_\tau-e$ NSI parameters were
fixed to a value close to the current upper bound from laboratory
experiments, $\epsilon^L_{\tau\tau}=\epsilon^R_{\tau\tau}=0.5$.

\begin{table*}
\begin{center}
\caption{Same as Table \ref{tab:eps_e} when both the $\nu_e-e$ and
$\nu_\tau-e$ NSI are non-zero. In all the cases we have fixed
$\epsilon^L_{\tau\tau}=\epsilon^R_{\tau\tau}=0.5$.}
\begin{tabular}{ccccccc}
\hline
$\epsilon^L_{ee}$ & $\epsilon^R_{ee}$ &
$z_{\rm fin}$ &
$\delta_e(\%)$ &
$\delta_{\mu,\tau}(\%)$ &
$N_{\rm eff}$  & $\Delta Y_p$\\ \hline
 0.1  &  0.5  & 1.3963 & 1.30 & 0.95 & 3.073 & 3.6${\times} 10^{-4}$\\
-1.0  &  1.0  & 1.3960 & 1.41 & 0.99 & 3.077 & 3.7${\times} 10^{-4}$\\
-1.5  & -1.0  & 1.3963 & 1.28 & 0.94 & 3.073 & 3.4${\times} 10^{-4}$\\
-1.0  &  0    & 1.3977 & 0.54 & 0.66 & 3.047 & 2.6${\times} 10^{-4}$\\
\hline\label{tab:eps_e_tau}
\end{tabular}
\end{center}
\end{table*}

\begin{table*}
\begin{center}
\caption{Same as Table \ref{tab:eps_e} when
the flavor-changing NSI parameters $\epsilon^{L,R}_{e\tau}$
are non-zero. The last two cases correspond to the results
for all NSI parameters non-zero with values given by
the bounds in Eqs.\ (\ref{bounds_Barranco}).}
\begin{tabular}{cccccccccccc}
\hline
$\epsilon^L_{ee}$ & $\epsilon^R_{ee}$ &
$\epsilon^L_{\tau\tau}$ & $\epsilon^R_{\tau\tau}$ &
$\epsilon^L_{e\tau}$ & $\epsilon^R_{e\tau}$ &
$z_{\rm fin}$ &
$\delta_e(\%)$ &
$\delta_{\mu,\tau}(\%)$ &
$N_{\rm eff}$  & $\Delta Y_p$\\ \hline
0 & 0 & 0 & 0 & 0.4 & 0.7 &
1.3960 & 1.32 & 1.03 & 3.077 & 3.9${\times} 10^{-4}$\\
0 & 0 & 0 & 0 & 0.85 & 0.38 &
1.3956 & 1.46 & 1.16 & 3.085 & 4.4${\times} 10^{-4}$\\
0.12 & 0.15 & 0.5 & 0.5 & 0.85 & 0.38 &
1.3949 & 1.68 & 1.38 & 3.098 & 5.0${\times} 10^{-4}$\\
-0.61 & -1.58 & -0.5 & -0.5 & -0.85 & -0.38 &
1.3948 & 1.69 & 1.41 & 3.100 & 5.1${\times} 10^{-4}$\\
0.12 & -1.58 & -0.5 & 0.5 & -0.85 & 0.38 &
1.3937 & 2.21 & 1.66 & 3.120 & 6.0${\times} 10^{-4}$\\
\hline\label{tab:offdiag}
\end{tabular}
\end{center}
\end{table*}

(4) Cases where the flavor-changing NSI parameters $\epsilon^{L,R}_{e
  \tau}$ are non-zero, that for simplicity we consider real. We
  consider both cases with vanishing diagonal NSI and with all possible
  NSI parameters with the largest allowed values according to Eqs.\
  (\ref{bounds_Barranco}).

The last column ($\Delta Y_p$) in Tables 1-4 reports the variation
in the mass yield of $^4$He synthesized during BBN, which predicts
typically $Y_p\simeq 0.248$ for the baryon density
$\omega_b=0.0223\pm0.0008$ favoured by CMB anisotropy
studies~\cite{Spergel:2006hy}. Neutrinos affect the outcome of BBN
in a flavor independent way through their contribution to the
radiation energy density (change in $N_{\rm eff}$). In addition, the
role played by electron neutrinos and antineutrinos in the weak
reactions that convert neutrons and protons is very important for
fixing the primordial production of $^4$He, which is thus the
nucleus mostly sensitive to non-standard physics in the neutrino
sector. To predict quantitatively the change $\Delta Y_p$ induced by
NSI, we take exactly into account the modified thermodinamical
quantities, while treating perturbatively the change in the weak
rates. This is justified given the relatively small effects we are
considering.  We closely follow the treatment given
in~\cite{Mangano:2005cc}, to which we refer the reader for further
details.

The results are generally in agreement with the approximate ones
predicted in Sec.\ \ref{sec:decoupling} (see in particular
Fig.~\ref{eps_neff}), apart for minor differences due to the improved
treatment. Also, as one naively expects, large $\epsilon^{L,R}_{ee}$
enhance mostly the distortions in the electronic flavor, while large
$\epsilon^{L,R}_{\tau\tau}$ lead to larger distortions in the tau
neutrino spectra. However, the differences between the spectra of the
three neutrino flavors are reduced by the partial re-shuffling of the
entropy transfer due to the effects of neutrino oscillations.

In general, we find that the presence of non-zero NSI enhances the
transfer of entropy from $e^\pm$ to neutrinos, leading to values up to
$N_{\rm eff}\simeq 3.12$ when all NSI parameters are close to the
boundaries of the allowed regions discussed in Sec.\ \ref{NSIsummary},
which is almost three times the departure of $N_{\rm eff}$ from $3$
that already exists for SM weak interactions (the first row in
Table 1). We also obtained some significant departures of $N_{\rm
eff}$ from the standard value, such as $N_{\rm eff}\simeq 3.36$ for
$\epsilon^L_{ee}=\epsilon^R_{ee}=4$, but only for NSI parameters well
beyond the allowed regions.  On the other hand, for particular choices
of the diagonal NSI parameters that minimize the $\nu-e$ interaction
we find a reduction in the final value of the distortions or $N_{\rm
eff}$ with respect to the SM value, as expected from our discussion in
Sec.\ \ref{decT_NSI}.

Finally, we mentioned in Sec.\ \ref{NSIsummary} that neutrino-neutrino
interactions are poorly constrained. In the case of interactions
involving left-handed neutrinos only, they can still be twice as
effective as in the SM, as reported in \cite{Bilenky:1999dn}. Since
enhanced neutrino-neutrino interactions will redistribute the
distortions more efficiently, they can modify the results described in
this section. In order to check this, we doubled the intensity of the
contributions of the neutrino-neutrino interactions to the collision
integral and repeated the calculations for some of the cases
considered above.  We found that the results are basically unchanged,
confirming the unimportant role played by neutrino-neutrino collisions
in the process of neutrino decoupling.

\section{Conclusions}\label{conclus}

The process of relic neutrino decoupling in the early Universe is
sensitive to the strength of the interactions between neutrinos and
the plasma formed by electrons and positrons. If neutrinos were kept
in longer thermal contact with them than in the standard case, they
would share a larger fraction of the entropy release from $e^\pm$
annihilations.  This would affect the predicted characteristics of the
cosmic background of neutrinos (C$\nu$B), which in turn could modify
the late evolution of the Universe and the bounds on neutrino
properties from the analysis of cosmological observables.

In this paper we have considered how the decoupling of relic neutrinos
is modified in presence of non-standard neutral current
neutrino-electron interactions. First, we have provided a rough
estimate of the size of NSI couplings needed to affect in a relevant
way the properties of the C$\nu$B and of related observables. We find
that needed couplings are by far larger than the existing laboratory
bounds, as also confirmed by a semi-analytical solution of the
relevant kinetic equations.  Thus, NSI can only play a minor role in
shaping the C$\nu$B.  However, NSI might still contribute with
comparable or larger distortions in the neutrino spectra than
predicted in the SM.  In order to quantify these
deviations, we have performed fully numerical and momentum-dependent
calculations of the density-matrix equations relevant for neutrino
evolution in the early Universe, including the effects of flavor
neutrino oscillations.

Typically, we find that the enhancement in observables like $\neff$,
which measures the change in the radiation content, or the $^4$He
mass yield in the BBN can be up to three times larger than the ones found
in the standard case \cite{Mangano:2005cc}, for values of all NSI 
parameters close to the limits placed by
laboratory experiments. Nevertheless, even the variation of up to
$0.2\%$ that we found in the $^4$He abundance is yet too small
(about one order of magnitude) when compared to the observational
error on $Y_p$, which is unfortunately dominated by systematics, and
as such difficult to pin down in the near future. Instead, a value
of $\neff\simeq 3.12$, which is still possible within the present
parameter space for NSI couplings, might even be barely detectable.
Indeed, it has been shown that the CMB satellite PLANCK will soon
provide temperature and polarization data that will probably measure
the value of $\neff$ with an uncertainty of order
$\sigma(\neff)\simeq 0.2$ \cite{Bowen:2001in,Bashinsky:2003tk} (for
a previous, more optimistic forecast see \cite{Lopez:1998aq}), or
better when combined with data from a large galaxy redshift survey
such as SDSS \cite{Lesgourgues:2004ps}. Future CMB missions may
reach the sensitivity of $0.04-0.05$ needed to test the standard
scenario \cite{Bashinsky:2003tk}, and might achieve a 2$\,\sigma$
hint for $\neff\simeq 3.12$ and eventually test the effect of large
neutrino-electron NSI.

However, barring some extreme cases, we conclude that the prediction
of $\neff\simeq 3$ within a few \% is quite robust even when taking
into account neutrino-electron NSI of the four-fermion type. Thus,
their existence can not modify in a significant way the bounds on
neutrino properties from cosmological observables, in particular on
their masses, as recently reviewed in \cite{Lesgourgues:2006nd}.
Since it is likely that future experiments may narrow further the
allowed range of NSI couplings, their role in a cosmological context
will be even smaller, at most a sub-leading correction to the
standard prediction of the C$\nu$B properties.  Turning the argument
around, however, one can conclude that a significant deviation of
$\neff$ from 3 may require major revisions of the
cosmological model, like the introduction of new relativistic relics
and/or of an exotic thermal history, or both (see e.g.\
\cite{Serpico:2004nm,Cuoco:2005qr}).

\section*{Acknowledgments}

We thank Omar Miranda and C\'elio A.\ de Moura for fruitful
discussions. This work was supported by a Spanish-Italian AI, the
Spanish grants FPA2005-01269 and GV/05/017 of Generalitat Valenciana,
as well as a MEC-INFN agreement. SP was supported by a Ram\'{o}n y
Cajal contract of MEC.  PS acknowledges the support by the Deut\-sche
For\-schungs\-ge\-mein\-schaft under grant SFB 375 and by the European
Network of Theoretical Astroparticle Physics ILIAS/N6 under contract
number RII3-CT-2004-506222.

\end{document}